\documentclass[aps,twocolumn,pra,showpacs]{revtex4}
\usepackage[dvips]{epsfig}

\date{\today}

\begin{document}

\title{How particle collisions increase the rate of accretion \\ from the cosmological background onto primordial black holes \\ in braneworld cosmology}

\author{V.~V.~Tikhomirov \footnote {vvtikh@mail.ru} and Y.~A.~Tsalkou \footnote {skytec@mail.ru}}
\address
{Institute for Nuclear Problems, Belarussian State University,
Bobruiskaya 11, 220050 Minsk, Belarus}

\draft
\begin{abstract}
It is shown that, contrary to the widespread opinion, particle
collisions considerably increase accretion rate from the
cosmological background onto 5D primordial black holes formed
during the high-energy phase of the Randall-Sundrum Type II
braneworld scenario. Increase of accretion rate leads to much
tighter constraints on initial primordial black hole mass fraction
imposed by the critical density limit and measurements of
high-energy diffuse photon background and antiproton excess.
\end{abstract}
 \maketitle

It is well known that primordial black holes (PBHs) open up a way
to obtain information on earliest era of the Universe
\cite{zel,haw}. Recent works \cite{gcl1,maj,gcl2,gcl3,sen1,sen2}
show that PBHs can also be a probe of braneworld cosmologies. The
Hawking temperature drop \cite{gcl1} and considerable PBH mass
growth due to radiation dominated background accretion occur
\cite{maj} in the simplest braneworld cosmology of the
Randall-Sundrum Type II (RS2). These modifications of PBH
properties  provide new constraints on the initial PBH mass
fraction \cite{gcl3,sen1,sen2}. PBH mass growth is exponentially
sensitive to the accretion efficiency. However, authors of
\cite{gcl3,sen1,sen2} considered the accretion efficiency as a
free parameter. In fact, they started with $F$ equals 1 in the
case of collisionless radiative background and assumed that it can
only {\it decrease} when relativistic particle scattering length
is small. They also predicted that a sufficiently low value of $F$
leads to weaker constraints than those obtained in the standard 4D
cosmology. It also allow the restriction of the RS2 curvature
radius $l$ from below. Meanwhile, according to the theory of
spherical accretion of nonrelativistic particles \cite{sha}, the
hydrodynamic accretion theory \cite{zel2} should be used instead
of collisionless accretion theory \cite{bon,mic} for small
scattering length. Since {\it"the presence of collisions between
particles restricts tangential motion and funnels particles
effectively in the radial direction for efficient capture"}
\cite{sha}, the hydrodynamic accretion rate substantially exceeds
the collisionless one in the case of nonrelativistic particles.
The aim of the present paper is to demonstrate that collisions of
relativistic particles analogously {\it increase} the accretion
rate of the radiation dominated background and lead to the new,
much tighter, constraints.

Let us remind the reader that in the RS2 model our Universe is a
4D Lorentz metric hypersurface called a {\it brane}, embedded in a
$Z_2$ symmetric 5D anti-de Sitter spacetime called the {\it bulk},
characterized by the curvature radius $l$, presently bound
\cite{long} by condition $l \leq 0.2 mm$. A  black hole (BH)
behaves as an essentially 5D object described by the metric
\cite{gcl1,maj,gcl2,gcl3,sen1,sen2}
\begin {equation}
ds^2 = -\left(1-r_{BH}^2/r^2\right)  dt^2 +
\frac{1}{\left(1-r_{BH}^2/r^2\right)}dr^2 + r^2 d\Omega^2_2,
\end{equation}
where $d\Omega^2_2$ is the line element on a unit 2-sphere, if its
modified Schwarzchild radius
\begin {equation}
r_{BH} = \sqrt{\frac{8}{3 \pi}} \sqrt{\frac{l}{l_4}}
\sqrt{\frac{M}{M_4}} l_4
\end{equation}
is much smaller than $l$. We use the units where $c = G =1$ and
introduce 4D Planck scale quantities $l_4$, $M_4$ and $t_4$ for
convenience. The high energy brane regime is characterized
\cite{gcl1} by scale factor, energy density, horizon mass, and
duration
\begin {eqnarray}
 \nonumber
a=a_h  \left(\frac{t}{t_h}\right)^{1/2}, \qquad \rho = \frac{3}{32
\pi} \frac{M_4^2}{t_c \; t}, \\ M_h =8 M_4^2\frac{t^2}{t_c},
\qquad t_c = t_4 \frac{l}{2 l_4},
\end{eqnarray}
where $t$ is cosmic time, and $t_h < t_c$. It is usually assumed
\cite{zel,haw,gcl1,maj,gcl2,gcl3,sen1,sen2} that at formation time
$t_i$ the PBH mass is
\begin {equation}
M_i = f M_h(t_i).
\end{equation}
The condition $r_{BH} \ll l$ is fulfilled if $f \sim 1$ and $t_i <
t_c$.

The accretion rate
\begin {equation}
\frac{dM}{dt} =  F 4 \pi r^2_{BH} \; \rho(t)
\end{equation}
and energy density (3) lead to exponential BH mass growth
\begin {equation}
M(t) = M_i \left(\frac{t}{t_i}\right)^{2 F/\pi}
\end{equation}
at $t < t_c$ \cite{maj}. In fact, Eq. (5) is justified \cite{gcl2}
only in the case of collisionless relativistic particles when
accretion efficiency $F$ equals 1. However, a simple comparison of
the scattering length $l_{sc}$ with the radius (2) demonstrates
that the cosmological background cannot be considered as a
collisionless particle medium in the case of PBH masses most
relevant for observational constraints. In fact, the opposite
condition $l_{sc} \ll r_{BH}$ is true, in general, for such PBHs.
It allows one to treat the accreting cosmological background as a
continuous medium. Therefore, to obtain $F$ one should evaluate
the accretion rate (5) in hydrodynamic approximation.

Since radiation dominated background is characterized by an
indefinite particle number, the momentum conservation equation
\begin {equation}
(\rho+P)u_{i;k}u^k = -P_{,i}- u_i P_{,k}u^k,
\end{equation}
should be used \cite{lan} together with the entropy conservation
equation
\begin {equation}
(\sigma u^k)_{;k} = 0.
\end{equation}
Here comma and semicolon denote usual and covariant derivatives;
$\rho$, $\sigma$, $P$ and $u^i$ are energy and entropy densities,
pressure and 4-velocity of accretion flow, respectively. In the
metric (1) adiabatic accretion equations (7) and (8) take the form
\begin {equation}
uu^\prime + \frac{r_{BH}^2 }{r^3} +
\frac{P^{\:\prime}}{\rho+P}\left(
1+u^2-\frac{r_{BH}^2}{r^2}\right)=0
\end{equation}
and
\begin {equation}
\frac{\sigma'}{\sigma} + \frac{u'}{u} + \frac{2}{r}=0
\end{equation}
where $u = u^1$ is the radial component of 4-velocity and primes
denote the derivative with respect to $r$. The radiation dominated
background is described by equations \cite{kla}
\begin {equation}
\rho(T) = 3 P(T) = \frac{\pi^2}{30}g_{eff} T^4
\end{equation}
and
\begin {equation}
\sigma(T) = \frac{2\pi^2}{45}g_{eff} T^3,
\end{equation}
where $g_{eff}$ is the number of relativistic particle species at
temperature $T$. The value $g_{eff} \simeq 100$ will be used below
because $T \geq 1 TeV$ at $t < t_c$. Equations (11) and (12) allow
to rewrite Eqs. (9) and (10) in terms of temperature
\begin {equation}
uu^\prime + \frac{r_{BH}^2 }{r^3} + \frac{T^{\:\prime}}{T}\left(
1+u^2-\frac{r_{BH}^2}{r^2}\right)=0,
\end{equation}
\begin {equation}
3\frac{T'}{T} + \frac{u'}{u} + \frac{2}{r}=0.
\end{equation}
Solving Eq. (13) and (14) for $u'$ and $T'$, one obtains (compare
with \cite{sha})
\begin {equation}
u' = \frac{D_1}{D}, \qquad T' = -\frac{D_2}{D},
\end{equation}
where
\begin {equation}
D = \frac{2u^2-1 + r_{BH}^2/r^2}{u T},
\end{equation}
\begin {equation}
D_1 = \frac{2u^2 + 2 - 5 r_{BH}^2/r^2}{r T}
\end{equation}
and
\begin {equation}
D_2 = \frac{2u^2 - r_{BH}^2/r^2}{r u}.
\end{equation}
To avoid singularities in the accretion flow, the solution of Eqs.
(15) must pass through a "critical" or "sound" point $r =r_s$
where $D=D_1=D_2=0$. Solving equations $D=D_1=D_2=0$, one obtains
\begin {equation}
r_s= \sqrt{2} \; r_{BH}
\end{equation}
and
\begin {equation}
u_s = u(r_s) =\frac{1}{2}
\end{equation}
in light speed units.

Integration of Eq. (13) leads to a conservation law
\begin {equation}
\left( 1+u^2-\frac{r_{BH}^2}{r^2}\right) T^2= const = T^2_b;
\end{equation}
where $T_b$ is the background temperature defined by expressions
(3) and (11) for energy density \cite{gcl1}. Substituting Eqs.
(19) and (20) into Eq. (21) one obtains
\begin {equation}
T_s = T(r_s) = \frac{2}{\sqrt{3}} T_b.
\end{equation}

Direct integration of Eq. (10) shows that the term $r^2 \sigma(r)
u(r)$ does not depend on $r$, while Eqs. (19), (20) and (22) allow
to evaluate its value
\begin {equation}
r^2 \sigma(r) u(r) = r_s^2 \sigma(r_s) u(r_s).
\end{equation}
Applying the latter equation sufficiently far from BH, where $T(r)
\rightarrow T_b$, and using Eqs. (12), (19), (20), and (22), one
obtains
\begin {equation}
u(r) \simeq \frac{8}{3\sqrt{3}} \times \frac{r_{BH}^2}{r^2},
\qquad r \gg r_{BH}.
\end{equation}
Substituting $r \gg r_{BH}$, $T(r) \simeq T_b$, $\rho(r) \simeq
\rho(T_b) \equiv \rho(t)$ and Eq. (24) into the formula
\begin {equation}
\frac{d M}{dt} = 4 \pi r^2 \rho(r) u(r)
\end{equation}
we finally come to the accretion rate (5) with the accretion
efficiency
\begin {equation}
F = \frac{8}{3\sqrt{3}} \simeq 1.54
\end{equation}
describing radiation dominated background accretion in
hydrodynamic approximation applicable in the limit of frequent
particle collisions. Since (26) exceeds 1, particle collisions, in
fact, increase both accretion rate (5) and PBH mass growth
coefficient (6).

Let us now demonstrate that continuous approximation is indeed
valid for the most of the high-energy phase. Since the
characteristic space scale (19) of accretion flow is determined by
$r_{BH}$, one should compare $r_{BH}$ with the scattering length
$l_{sc} = l_{sc}(t) \simeq 1/\sigma n$. Here $n = n(T_b) \sim
\rho(T_b)/3 T_b$ is the particle number density and $\sigma =
\sigma(T_b) \simeq \alpha^2/T_b^2$ is the scattering cross section
in which $\alpha = \alpha(T_b)$ is the interaction constant.

Let us remind the reader that a substantial mass growth will be
experienced by PBHs with initial masses
\begin {equation}
M_{min} \leq M_i < M_{max}.
\end{equation}
The lower limit
\begin {equation}
M_{min} = 2 \cdot 10^6 \left(\frac{l}{l_4}\right)^{-1/3} M_4
\end{equation}
follows \cite{gcl1} from the absence of excessive large-angle
microwave anisotropy while the upper one
\begin {equation}
M_{max} \simeq \left(\frac{l}{l_4}\right) M_4
\end{equation}
follows from both the condition $r_{BH} = l$ and Eq. (4) at
$t=t_c$. Note that BHs with larger masses and radii are the usual
4D BHs which do not experience considerable accretion mass growth
\cite{maj,gcl2}.

For simplicity we will completely neglect scattering taking $F=1$
if $r_{BH} < l_{sc}$ and treat the radiation dominated background
as a continuous medium taking $F \simeq 1.54$ if $r_{BH} >
l_{sc}$. Let us introduce the initial PBH masses $M_{i1}$ and
$M_{i2}$ for which the condition $r_{BH} = l_{sc}$ fulfills,
respectively, at the final moment $t = t_c$ of the high-energy
phase and at the moment $t = t_i$ of the PBH formation. These
masses allow us to separate the mass intervals corresponding to
different scenarios of PBH mass growth (see Fig. 1), namely, the
interval $M_{min} < M_i < M_{i1}$ where $l_{sc} > r_{BH}$ and
$F=1$, the interval $M_{i2} < M_i < M_{max}$ where $l_{sc} <
r_{BH}$ and $F \simeq 1.54$ and the mixed interval $M_{i1} < M_i <
M_{i2}$ for which a moment $t_i < t_* < t_c$ exists at which
$r_{BH}(t_*) = l_{sc}(t_*)$ and one, correspondingly, has $r_{BH}
< l_{sc}$, $F=1$ at $t_i < t < t_*$ and $r_{BH} > l_{sc}$,
$F\simeq 1.54$ at $t_* < t < t_c$. Equation (6) can now be
directly applied to each interval with constant $F$ giving a
composite formula
\begin{equation}
\frac{M(t_c)}{M_i} = \cases{ \left(\frac{t_c}{t_i}\right)^{2/
\pi}, \hspace{22mm} M_{min} < M_i < M_{i1}; \cr
\left(\frac{t_*}{t_i}\right)^{2/\pi}
\left(\frac{t_c}{t_*}\right)^{16/(3 \sqrt{3} \pi)}, \hspace{5mm}
M_{i1} < M_i < M_{i2}; \cr \left(\frac{t_c}{t_i}\right)^{16/(3
\sqrt{3} \pi)}, \hspace{17mm} M_{i2} < M_i < M_{max}}
\end{equation}
for the coefficient of accretion mass growth during the
high-energy phase. The solid lines on Fig. 1 show the coefficient
of accretion mass growth as a function of $M_i$ for curvature
radii $l/l_4 = 10^{21}, 10^{26}$, and $10^{31}$ and $f=0.1$. Long
dashed and short dashed lines correspond, respectively, to the
values $F=1$ and $F=0.5$ used in \cite{gcl3,sen1,sen2}. From Fig.
1 one can see that Eq. (30) predicts considerably larger mass
growth than Eq. (6) with $F = 1$ and 0.5. One can also see that
the region $M_{i2} < M_i < M_{max}$ in which Eq. (30) reduces to
Eq. (6) with (26) is quite wide.

\begin{figure}[!ht]
\centering \psfull
    \epsfig{file=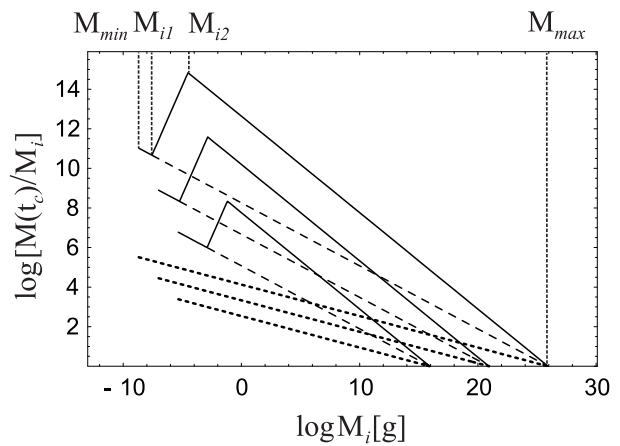,width=8cm}
    \caption{Accretion mass growth coefficient vs the initial PBH mass
    for choices $l/l_4 =10^{21}, 10^{26}$ and $10^{31}$ (from bottom
    to top). Solid lines correspond to the PBH mass growth
    coefficient (30), while long dashed and short dashed ones are obtained
    from Eq. (6) with fixed accretion efficiencies $F=1$ and $F=0.5$,
respectively.}
\end{figure}

Let us keep in mind that the curvature radius is experimentally
bound from above by the value $0.2 mm \sim 10^{31}l_4$
\cite{long}, while its lower limit is determined by the choice of
5D BH evaporation time \cite{gcl1,maj,gcl2,gcl3,sen1,sen2}
\begin{equation}
\frac{t_{evap}}{t_4}  \simeq \tilde{g} ^{-1} \frac{l}{l_4}
\left(\frac{M(t_c)}{M_4}\right)^2,
\end{equation}
where $\tilde{g} =0.023$. Since the most direct ways of PBH search
are connected with PBHs completing their evaporation now or at a
future time, we limit $l$ by $l_{min} = 10^{20}l_4$ \cite{gcl1}
which follows from (31) at $t_{evap} = t_0 = 13.7 Gyr$ and
$M(t_c)=M_{max}$. Note that at $l/l_4 < 10^{20}$, one will not see
any difference from the 4D case.

The accretion efficiency (26) allows us to tighten the constraints
imposed on the initial PBH mass fraction \cite{gcl3,sen1,sen2}
\begin{equation}
\alpha_i = \alpha_i(M_i) =
\frac{\rho_{PBH,M_i}(t_i)}{\rho_{rad}(t_i)},
\end{equation}
where $\rho_{PBH,M_i}(t_i)$ is the mass density of PBHs with
formation mass on the order of $M_i$ and $\rho_{rad}(t_i)$ is the
radiation energy density at the moment $t_i$ of such PBH
formation. The mass $M_i$ should be evaluated substituting the
mass $M(t_c)$ found from Eq. (31) at given $l$ and $t_{evap} =t_0$
into Eq. (30). The PBH mass fraction grows with time starting from
the initial value (32) proportionally to $a(t)M(t)$ at $t_i < t <
t_c$ and proportionally to $a(t)$ at $t_c < t < t_0$ \cite{gcl3}.
Describing the PBH mass growth by Eq. (30) and restricting the
present PBH mass density, by 30\% of the critical density one
obtains new constraints on $\alpha_i$ presented by the solid line
in Fig. 2. The latter are much tighter than both the 4D constraint
$\alpha_i < f^{-1/2} 10^{-18}$ and the old 5D constraints
\cite{gcl3} illustrated by long dashed and short dashed lines
evaluated using Eq. (6) with $F=1$ and $F=0.5$, respectively. The
dotted lines both in Fig. 2 and in Figs. 3, 4 below are evaluated
using the third line of Eq. (30) in order to illustrate the degree
of violation of hydrodynamic approximation in the region $l/l_4
\geq 10^{30}$ where the accretion actually begins in a
collisionless regime and only later switches over into a
hydrodynamical one being, thus, adequately described by the second
line of Eq. (30).

\begin{figure}[!ht]
\centering \psfull
    \epsfig{file=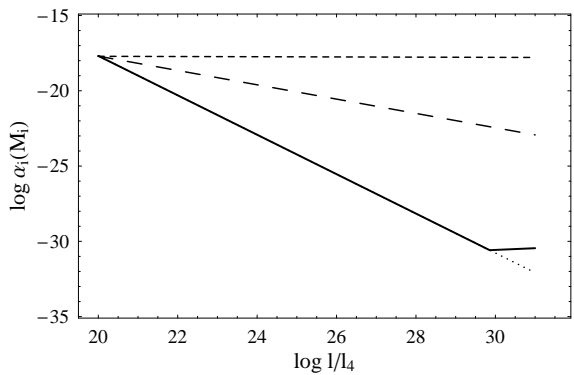,width=8cm}
    \caption{Constraints on initial PBH mass fraction imposed by the total
matter density. Dotted, long dashed and short dashed lines are
obtained using PBH mass growth coefficient (6) with $F=1.54, 1$
and $0.5$, respectively.}
\end{figure}

Equation (30) also allows to revise the constraints following from
the measured high-energy diffuse photon spectrum \cite{sen1} and
antiproton excess \cite{sen2}. Fortunately, it is sufficient to
consider the PBHs intensively evaporating at present in both cases
\cite{sen1,sen2}. The mass $M^*(l) = M(t_c) \propto l^{-1/2}$ and
Hawking temperature $T^*_{BH} = 1/2 \pi r_{BH} \propto
\sqrt{M^*(l) l} \propto l^{-1/4}$ of such PBHs can be found
substituting $t_{evap} =t_0$ into Eq. (31) and then $M^*(l)$ from
Eq. (31) into Eq. (2). The Hawking temperature determines the
photon energy $E \sim b T^*_{BH}$ with $b \simeq 5$ \cite{gcl3}
increasing proportionally to $l^{-1/4}$ from 250 keV to 150 MeV
with the curvature radius variation from $l/l_4 = 10^{31}$ to
$10^{20}$ \cite{gcl1}. The diffuse photon spectrum can be
estimated \cite{gcl3,sen1} by the formula
\begin{equation}
I(E)= \frac{c}{4 \pi} \frac{M^*}{E} n(t_0, M_i(M^*))
\end{equation}
where $c$ is the speed of light and
\begin{eqnarray}
 \nonumber
n(t_0, M_i(M^*)) = \\ \frac{3 \alpha_i(M_i)}{2^{17/4} \pi
l_4^{9/4}} \frac{a^3(t_{eq})}{a^3(t_0)} f^{1/8} l^{-3/8}
t_{eq}^{-3/2} M_i^{-9/8}(M^*)
\end{eqnarray}
is the present number density of PBHs with mass of the order
$M^*$, where $t_{eq}=72.6 kyr$ \cite{sen1,sen2} is the time of
matter-radiation equality and the mass $M_i$ should be found by
substituting $M(t_c) = M^*$ into Eq. (30). In particular, one can
find that initial masses of PBHs completing their evaporation at
present are equal to $M_i = 10^{-4}$ and $10^3$g for $l/l_4 =
10^{31}$ and $10^{20}$, respectively. Comparing (33) with the
observational data collected in \cite{sen1}, one obtains new
constraints presented in Fig. 3 by a solid line. These constraints
are much tighter than both the 4D constraint $\alpha_i < 10^{-27}$
and the old 5D ones, presented by dashed lines.

\begin{figure}[!ht]
\centering \psfull
    \epsfig{file=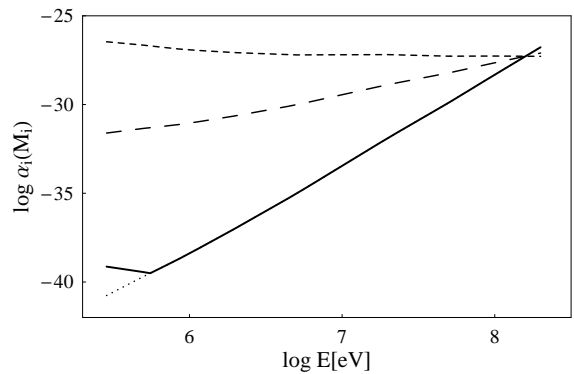,width=8cm}
    \caption{Constraints on initial PBH mass fraction imposed by the
high-energy diffuse photon background. Dotted, long dashed and
short dashed lines are obtained using Eq. (6) with $F=1.54, 1$ and
$0.5$, respectively.}
\end{figure}

The antiproton flux from brane PBHs can be schematically written
in the form
\begin{equation}
\Phi_{\bar{p}} \propto \frac{dM_i}{dM^*} \frac{M_{GeV}^2}{M_i M^*}
{\it G} n(t_0, M_i(M^*)),
\end{equation}
where $M_{GeV} = M_{GeV}(l)$ is the mass of a PBH with Hawking
temperature $T_{BH} \sim 1 GeV$ and the factor $G \sim 10^{5}$
represents the galactic matter density enhancement at the solar
neighborhood. Equation (35) differs from equations of ref.
\cite{sen2}, combined together in a similar way, solely by the
relation (30) of $M^* = M(t_c)$ with $M_i$. This "small"
difference, however, allows us to considerably restrict the "most
important result" of Ref. \cite{sen2} concerning the $\bar{p}$
flux dependence on $l$. Indeed, using the PBH mass growth
coefficient (6), one obtains \cite{sen2} from (35) that
$\Phi_{\bar{p}} \propto l^p$ with exponent
\begin{equation}
p = \frac{40F-13 \pi}{16(\pi - F)}.
\end{equation}
Since $p < 0$ at $F\leq 1$ it was argued in \cite{sen2} that the
$\bar{p}$ flux is a {\it decreasing} function of $l$ (see both
dashed lines in Fig. 4) and should be used to set a lower bound on
$l$. However the new value $F \simeq 1.54$ leads to the positive
exponent $p \simeq 0.81$ corresponding to {\it increasing}
$\bar{p}$ flux (35) and decreasing constraints on $G \alpha_i$
(solid line in Fig. 4) up to $l/l_4 \simeq 10^{30}$, where the
hydrodynamic approximation is violated. These constraints are
again much tighter than both the 4D constraint $\alpha_i <
10^{-27}$ and the old 5D ones, presented by dashed lines.

\begin{figure}[!ht]
\centering \psfull
    \epsfig{file=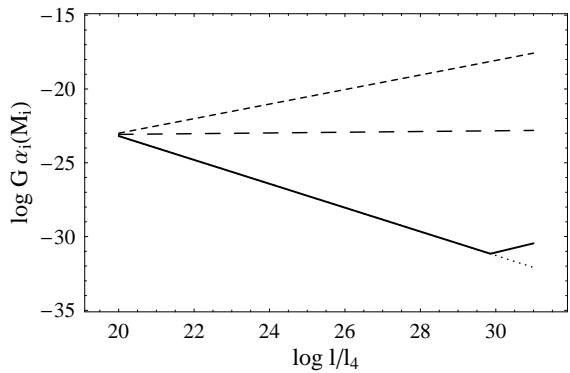,width=8cm}
    \caption{Constraints on initial PBH mass fraction imposed by the
antiproton excess. Dotted, long dashed and short dashed lines are
obtained using Eq. (6) with $F=1.54, 1$ and $0.5$, respectively.}
\end{figure}

Because of violation of hydrodynamic approximation at $l/l_4 >
10^{30}$ and in the corresponding photon energy region $E \leq 500
keV$ all the considered constraints reach their maxima at $l/l_4
\simeq 10^{30}$. Figures 2-4 demonstrate that the tightest
constraints following from the critical density limit and
measurements of the high-energy diffuse photon background and
antiproton excess at $l/l_0 \simeq 10^{30}$ amount, respectively,
to $\alpha_i \simeq 10^{-31}, 10^{-39}$ and $10^{-36}$. Since the
tightness of all the constraints increases with the curvature
radius $l$, all of them can be used to set an upper bound on $l$
at $l/l_4 \leq 10^{30}$.

To summarize, particle collisions make hydrodynamic approximation
adequate to describe the accretion of the radiation dominated
background onto PBHs during most of the high-energy phase of the
RS2 braneworld scenario. Continuous background is characterized by
54\% higher accretion efficiency onto PBHs than that of
collisionless relativistic particles. This higher accretion
efficiency provides larger PBH mass growth during the high-energy
phase and makes it possible to obtain much tighter constraints on
initial mass fraction than those found in \cite{gcl3,sen1,sen2} as
well as than the constraint equal to $10^{-27}$ obtained in the
standard 4D cosmology. The tightness of the obtained constraints
mostly increases with curvature radius allowing to restrict it
from above.

The authors are grateful to the Ministry of Education and
Professor V. G. Baryshevsky for support of this work.

\end{document}